\documentclass{styles/svproc}
\usepackage{float}
\usepackage{amsmath}
\usepackage{graphicx}
\usepackage{subcaption}

\usepackage{url}

\begin{document}
\mainmatter              
\title{Fabrication of semi-cylindrical channels for one-dimensional fiber array edge couplers}
\titlerunning{Fiber arrays}  
%
\author{Sakthi Priya Amirtharaj\inst{1} \and Saandeep Peesapati\inst{2} \and Sirshi S Ram\inst{3}  \and Sivarama Krishnan\inst{3} \and Anil Prabhakar\inst{1}}

\authorrunning{Sakthi Priya Amirtharaj et al.} 
%

%
\institute{Department of Electrical Engineering, \\Indian Institute of Technology Madras, Chennai, India.\\
\email{anilpr@ee.iitm.ac.in}
\and
Department of Physics, \\Indian Institute of Science Education and Research, Mohali, India
\and
Department of Physics, \\Indian Institute of Technology Madras, Chennai, India.}

\maketitle              

\begin{abstract}

Photonic Integrated Circuits (PICs) are essential for high-speed and compact optoelectronic applications, yet efficient optical coupling to PIC remains a critical challenge, where minimizing insertion losses is crucial for high-performance applications. Fiber arrays are commonly used as edge couplers for PICs. In this work, we demonstrate the fabrication of semi-cylindrical channels on glass substrates using femtosecond laser micromachining for fiber arrays edge couplers. This method enables the formation of narrow, well-defined grooves in glass substrates with submicron accuracy, facilitating reliable fiber positioning for improved coupling efficiency. Experimental results demonstrate the feasibility of this approach for dense fiber arrays, with a narrow separation of $\sim$ 3 \textmu m.

\end{abstract}
\section{Introduction}
Photonic integrated circuits have emerged as a cornerstone of modern photonic technology, offering compact, scalable, and highly functional platforms for optoelectronic applications, including telecommunications, data processing, sensing, and quantum computing \cite{shekhar_roadmapping_2024,pelucchi_potential_2021}. By integrating multiple optical components—such as waveguides, modulators, and detectors—onto a single substrate, PICs enable high-speed operation and exhibit CMOS compatibility, facilitating seamless integration with semiconductor manufacturing processes. While silicon photonic elements have seen significant advancements \cite{thomson_roadmap_2016}, a major bottleneck in PICs is the optical interconnects, particularly the efficient coupling of light into the PIC. This challenge primarily arises from the size mismatch between the mode field diameters of the two components: the fiber core is approximately 10~\textmu m, while waveguides are on the order of hundreds of nanometers. Efficiently coupling external light into the submicron-scale waveguides of PICs is critical for their performance and scalability. Grating coupling and edge coupling are the most commonly used techniques for fiber-to-PIC coupling \cite{mu_edge_2020,marchetti_coupling_2019,carroll_photonic_2016}. Grating couplers offer higher alignment tolerances but suffer from high insertion losses and strong polarization dependence, making them unsuitable for fiber inputs that normally have unknown or unstable polarization. Edge coupling, by contrast, is more promising due to its high efficiency, broad bandwidth, and polarization independence \cite{mu_edge_2020}.

In edge coupling, fibers are positioned in the same plane as the waveguides, which have tapered edges with the narrow end placed near the fiber. Lensed fiber tips are often employed to focus light onto the waveguides, enhancing coupling efficiency. In many applications, multiple fiber channels must be coupled to the PIC. Fiber arrays, which consist of optical fibers arranged in an orderly manner to match the input/output positions of PIC waveguides, enable simultaneous coupling to multiple ports. V-grooves are engraved in silicon or glass substrates to hold the fibers in place at a pitch of 127~\textmu m or larger, matching the size of telecom fibers, with a lid secured on top for stability. Common techniques for fabricating V-grooves include wet etching \cite{lee_anisotropic_1969,uddin_materials_2009}, diamond-wheel machining \cite{xie_micro-grinding_2012,wang_study_2014}, and lithography-based approaches \cite{wu2017,stassen2020}. The fabrication tolerances of fiber arrays are much higher than those required for edge couplers \cite{carroll_photonic_2016}, and even today, achieving high-precision fiber arrays with narrow gaps of a few micrometers remains challenging, directly impacting their performance in PICs.

In this study, we explore a precise and time-efficient fabrication method for one-dimensional fiber arrays using femtosecond laser micromachining \cite{gattass2008femtosecond,della_valle_micromachining_2009}. This approach enables the fabrication of high-density and custom-designed grooves for fiber alignment on glass or silicon substrates \cite{gross2015}. The method provides submicron accuracy and flexibility in array geometry, making it particularly suitable for PIC applications \cite{zhou2018}. Utilizing ultrafast femtosecond pulses, this technique selectively ablates material while minimizing thermal diffusion and collateral damage, resulting in sharp, well-defined features. Here, we demonstrate the fabrication of fiber arrays with an extremely narrow gap between fibers (3~\textmu m) by etching U-shaped grooves into a glass substrate using femtosecond galvoscanning \cite{macernyte_femtosecond_2019}. We further characterize these fabricated grooves as channels for fiber arrays, showcasing their effectiveness for PIC integration.

\section{Fabrication of the fiber array edge couplers}
\begin{figure}[H]
    \centering
     \begin{subfigure}{0.38\textwidth}
        \includegraphics[height=0.18\textheight]{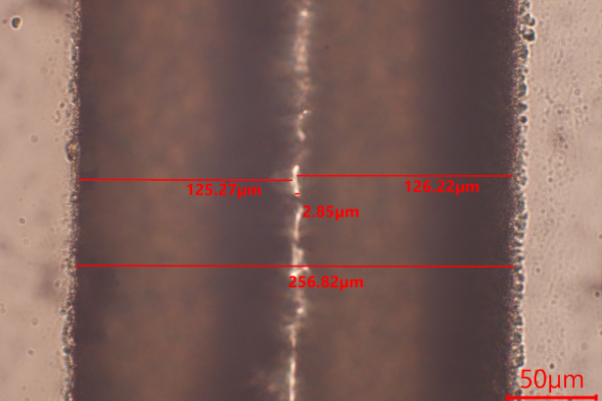}
        \caption{}
        \label{microimg_3Dprof}
    \end{subfigure}
    \hfill
    \begin{subfigure}{0.38\textwidth}
        \includegraphics[height=0.18\textheight]{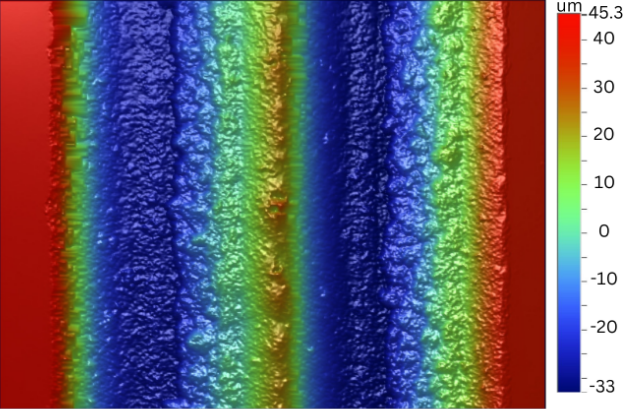}
        \caption{}
        \label{surface_profile}
    \end{subfigure}
    \caption{\small{(a) Microscopic image and (b) 3D surface profile of a pair of semi-cylindrical channels.}}
    \label{micro_surfaceprofile}
\end{figure}

U-shaped grooves are fabricated on a standard microscopic glass slide with a 127 \textmu m spatial separation. Compared to V-grooves, semi-cylindrical channels offer a smaller spatial pitch and a better fit for fibers. This is achieved using femtosecond galvoscanning, which microstructures channels with a depth of 62.5 \textmu m and a width of 125 \textmu m, creating U-shaped profiles on the glass substrate. It takes only 6 minutes to engrave 4 grooves of length 75~mm in a standard microscopic glass slide. Fig.~\ref{microimg_3Dprof} shows micro-channels having 3 \textmu m separation between the semi-cylindrical channels.

The laser system used is Coherent Monaco, delivering femtosecond pulses at 1030 nm, with a pulse width of 275 fs and energy per pulse of 40 \textmu J. The galvoscanner is equipped with an f-theta lens with a 160 mm focal length, providing a 20 \textmu m spot size. The channels are created by ablating rectangular hatches with a 3 \textmu m pitch for maximum overlap. These hatches are machined at a scanning speed of 50 mm/s and a repetition rate of 25 kHz (65\% RF), ensuring optimal depth, width, and a crack-free profile. The fabricated channels are then treated in ethanol for two minutes, and sonicated in isopropyl alcohol.  Fig.~\ref{surface_profile} shows the depth analysis performed using a 3D profilometer. The channels were separated by about 128 \textmu m as in the microscopic image and a U-shaped groove with a depth of about 77 \textmu m is ablated from the glass substrate.

\vspace{-2em}

\begin{figure}[H]
    \centering
     \begin{subfigure}{0.35\textwidth}
        \includegraphics[height=0.2\textheight]{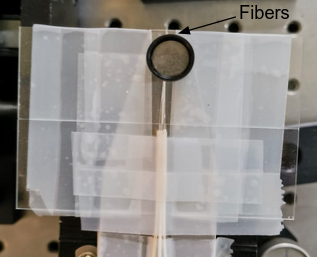}
        \caption{}
        \label{array_top}
    \end{subfigure}
    \hspace{2cm}
    \begin{subfigure}{0.35\textwidth}
    \centering
            \includegraphics[height=0.2\textheight]{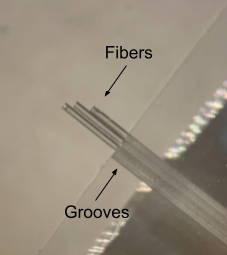}
        \caption{}
        \label{fin_in_grooves}
    \end{subfigure}
        \vspace{-1em}
    \caption{\small{(a) Image of three fibers placed in the grooves held with a magnetic clamp. (b) microscopic image of the three fibers alongside an empty groove.}}
    \label{fibers_topview}
\end{figure}

\vspace{-3em}
Single-mode telecom fibers are stripped to remove the outer jacket and protective layer, leaving only the fiber with a cladding diameter of 125 \textmu m. The tips of the fibers are cleaved. The fibers are then cleaned with isopropyl alcohol (IPA) and carefully placed in the grooves while being monitored under a microscope. A magnetic clamp holds them in place, as shown in Fig.\ref{fibers_topview}a. Alternatively, laser welding \cite{luo_femtosecond_2022} or epoxy can be used to secure the fibers to the substrate. Three fibers were placed in a substrate with four grooves. An empty groove, along with the three inserted fibers, is shown in Fig.\ref{fibers_topview}b.
\vspace{-1em}

\section{Beam profile characterization}
 The fiber array is characterized by measuring the beam profile of light emerging from each fiber using a knife-edge experiment. A 1550 nm laser input is fed to the fiber, and the output beam is measured with an InGaAs photodiode.
 

\subsection{Experimental setup}

\begin{figure}[H]
    \begin{subfigure}{0.45\textwidth}
        \centering
        \includegraphics[height=0.21\textheight]{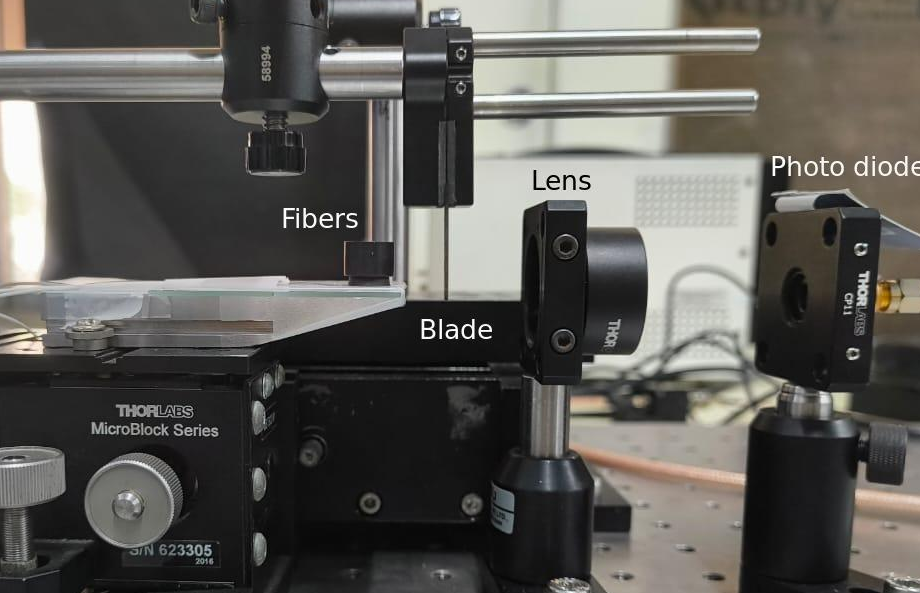}
        \caption{}
        \label{fig:KE_setup}
    \end{subfigure}
    \hfill
    \begin{subfigure}{0.45\textwidth}
    \includegraphics[height=0.21\textheight]{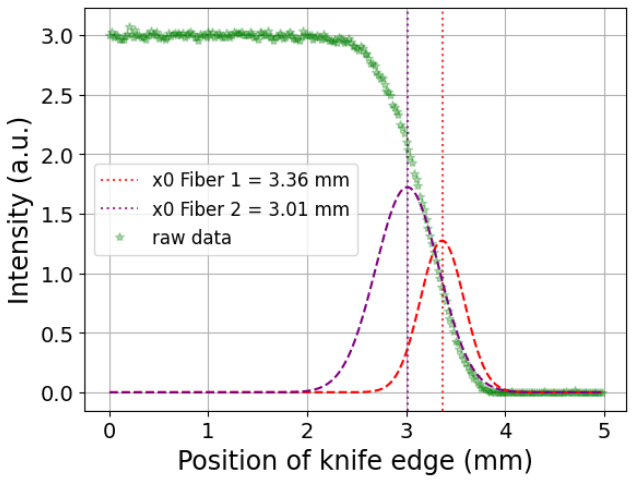}
    \caption{}
    \label{fig:R2L}
    \end{subfigure}
    \caption{\small{(a) Image of the knife-edge experimental setup. (b)The beam profile of the 2 fibers was measured using a knife-edge experiment. The raw data (green dots) is fitted with an error function and retrieved Gaussian peaks of the fibers are plotted (dotted lines) with the center at x0, indicated by the vertical lines}}
    \label{fig: beam profile}
\end{figure}

The knife-edge experiment is performed by placing an object with a sharp edge on the path of the laser beam and placing a photodiode after the knife-edge to measure the transmitted optical power~\cite{de2009measurement,gonzalez-cardel_gaussian_2013}. The knife-edge is gradually translated perpendicular to the beam propagation direction to intercept the laser beam. Fig.~\ref{fig:KE_setup} shows the experimental setup used to perform the knife-edge method. The intensity profile is measured using the InGaAs photodiode and is shown in Fig.~\ref{fig:R2L}. To clearly resolve the output from different fibers, light is fed into the first and third fibers in Fig.~\ref{fibers_topview}b. From the collected raw intensity data, the beam profile of the output from individual fibers is analytically estimated. Finally, we characterize the fiber alignment by calculating the divergence of light exiting the individual fibers. 

Assuming Gaussian distribution, the intensity profile of the laser beam is given by:

\begin{equation}
I(x, y) = I_0 \exp\left[ - \frac{2 (x - x_0)^2}{\omega_x^2} - \frac{2 (y - y_0)^2}{\omega_y^2} \right]
\end{equation}
where \( I_0 \) is the peak intensity; \( \omega_x \) and \( \omega_y \) are the radius of the beam where the intensity is \( 1/e^2 \) times $I_0$ in \( x \) and \( y \), respectively; and \( x_0 \) and \( y_0 \) are the values of \( x \) and \( y \) at the beam center.
The measured power in the photodiode for a given knife-edge position \( x \) is given by the integral of the Gaussian function and can be expressed in terms of error function \cite{gonzalez-cardel_gaussian_2013}:
\begin{equation}
P(x) = \frac{P_0}{2} \left[ 1 + \text{erf}\left( \frac{\sqrt{2}(x - x_0)}{\omega_0} \right) \right]
\label{eq: error function}
\end{equation}
where \( P_0 = \frac{\pi}{2} \omega_0^2 I_0 \) and \( \text{erf}(u) \) is the Gaussian error function defined as
\begin{equation}
\text{erf}(u) = \frac{2}{\sqrt{\pi}} \int_0^u e^{-t^2} \, dt.
\end{equation}
Once we measure the power output at the photodiode, we can then fit the data to the error function to retrieve the Gaussian parameters.

\subsection{Beam propogation characterisation}

The knife edge experiment is performed with two fibers in the grooves separated by a distance of $\sim 0.254\,$mm, while light also passed through them using a 1×2 splitter. The data from the experiment is fitted to a sum of two error functions expressed by Eq.~\ref{eq: error function}, corresponding to the individual fibers. A SciPy curve fit function is used to obtain the best fit of the data. The Gaussian profile of the individual fibers recovered from the data fit is shown as dotted lines in Fig.~\ref{fig: beam profile}. The difference in the intensity ratio of the two Gaussian peaks is attributed to the difference in the transmission of the two fiber-splitter paths.

Now, we set out to retrieve the 3D map of the beam propagation. To achieve this, the knife-edge experiment is performed across the beam at 10 different distances from the fibers. The beam profile of light exiting a fiber optic cable is largely determined by the mode structure of the light within the fiber, the core size, and the numerical aperture (NA) of the fiber. The numerical aperture (NA) of a fiber defines the acceptance angle and can also be used to calculate the divergence angle of the beam exiting the fiber. For single-mode fibers, the divergence can be approximated using the fiber’s numerical aperture
\begin{equation}
\theta = \sin^{-1}(\text{NA})
\end{equation}
For small angles (as typical in fiber optics), $\theta \approx \text{NA}$, and we can deduce the beam waist as a function of $z$ as
\begin{equation}
\omega(z) = \omega_0 + z \tan(\text{NA})
\label{eq:wz}
\end{equation}

\begin{figure}[H]

  \begin{subfigure}[b]{0.5\textwidth}
    \includegraphics[height=0.75\textwidth]{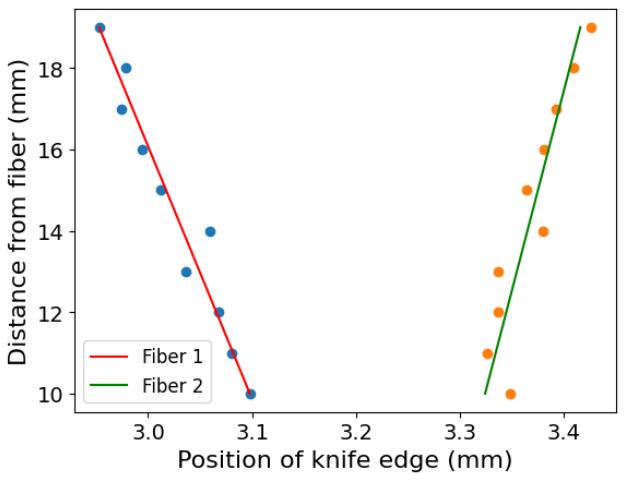}
    \caption{}
    \label{fig:peak_pos_best}
  \end{subfigure}
  \hspace{0.1cm}
  \begin{subfigure}[b]{0.5\textwidth}
    \includegraphics[height=0.75\textwidth]{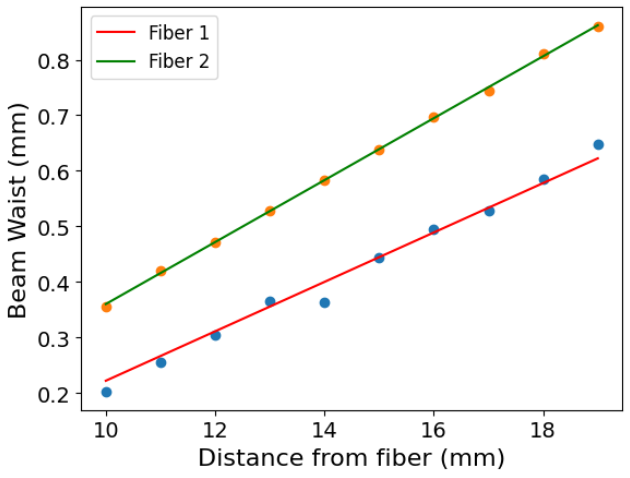}
    \caption{}
    \label{fig:beamWaist_fit}
  \end{subfigure}
  \begin{subfigure}[b]{0.5\textwidth}
    \includegraphics[height=0.75\textwidth]{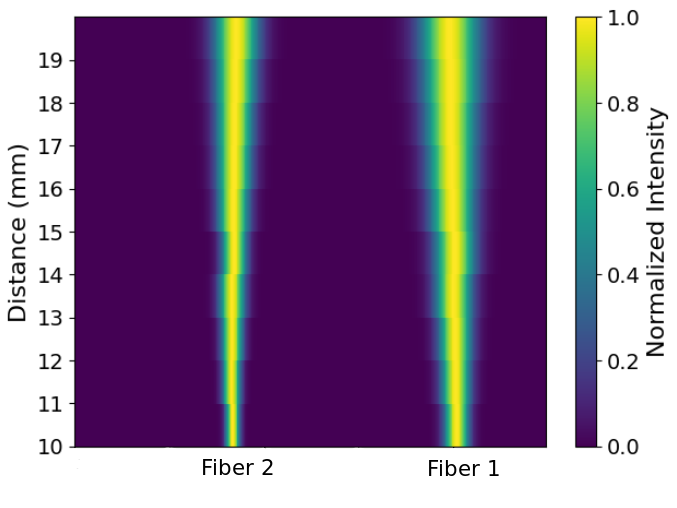}
    \caption{}
    \label{fig:3d_orig}
  \end{subfigure}
  \hspace{0.3cm}
  \begin{subfigure}[b]{0.5\textwidth}
    \includegraphics[height=0.75\textwidth]{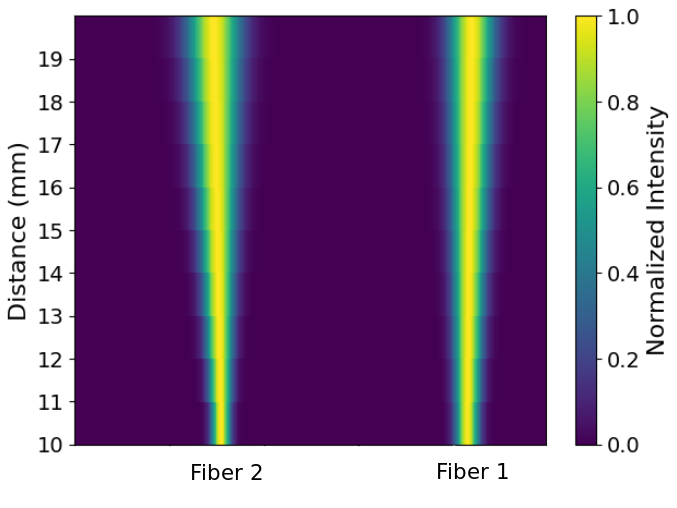}
    \caption{}
    \label{fig:3d_fit}
  \end{subfigure}
\caption{(a) Best fit curve of the peak positions of constructed Gaussians at various distances from the fiber tip. (b) Best fit curve of beam waists of constructed Gaussians at various distances from the fiber tip. (c) 3D color map plot of the constructed Gaussians for both fibers at various distances.}
\label{fig:Final_chact}
\end{figure}

From the results of the knife edge experiment, the Gaussian parameters of the fiber output are estimated for each distance $z$. The resulting beam waist as a function of $z$ is fit to Eq.~\ref{eq:wz}. Fig.~\ref{fig:beamWaist_fit} gives us an array of best-fit beam waist values for various z values. Similarly, as shown in Fig.~\ref{fig:peak_pos_best}, linear regression of the peak positions of the Gaussians of each fiber at various $z$ distances was also performed, and finally a three-dimensional beam profile is plotted with originally estimated Gaussian parameters (Fig.~\ref{fig:3d_orig}) and the values obtained from the data fitting of $\omega(z)$ and $P(z)$ (Fig.~\ref{fig:3d_fit}). The Gaussian data of the individual fibers have been normalized for better visualization.

From Fig.~\ref{fig:peak_pos_best}, we can calculate the misalignment angle between the fibers by considering the slopes of best-fit line equations from Fig.~\ref{fig:peak_pos_best} (fiber 1: $y = -0.02x + 3.26$ and fiber 2: $0.01x +  3.22$) by using the formula,
\begin{equation}
    \theta = \arctan \left[\frac{\mid m_1 - m_2\mid}{1 + m_1m_2}\right]
\end{equation} 
where $m_i$ is the slope of the fiber $i$.
The misalignment angle between the fibers separated by a groove (0.25mm) is estimated to be 1.51 $\pm$ 0.23 degrees. It is to be noted that the commercially available fiber arrays have a misalignment angle ranging from $1^o$ to $3^o$ \cite{schoonderbeek_design_2019,sylex,sqs_matrix_fiber_arrays}. This says that our method of fabrication stands as a reliable and effective method to fabricate these arrays.

\section{Conclusion and outlook}
We have demonstrated the fabrication of semi-cylindrical channels for fiber arrays through femtosecond galvoscanning. This method provides a reliable approach for fabricating crack-free and debris-free micro-scale channels on glass substrates, as observed using a surface profiler. We have developed a prototype fiber array with three fibers embedded in the grooves and have characterized their performance. Our fiber arrays perform on par with commercially available fiber arrays.

While this prototype device consists of three fibers, it can be extended to accommodate a larger number of fibers. Additionally, the femtosecond laser system offers versatility for other packaging-related steps, such as fiber cleaving \cite{mamun_quantifying_2021} and fiber-glass welding \cite{luo_femtosecond_2022}. For instance, it could be used to uniformly cut fiber edges and enable laser welding for bonding fiber-to-fiber or fiber-to-glass, eliminating the need for adhesives or mechanical clamps. This advancement paves the way for the complete packaging of fiber arrays using only the femtosecond laser system.

\section*{Author Contributions}
S.P. Amirtharaj designed the study, developed the experimental setup, performed the experiments, analyzed the data, and wrote the manuscript. S. Peesapati contributed to data collection and analysis, figure preparation, and participated in manuscript writing. S.S. Ram fabricated the devices and contributed to the manuscript section on fabrication. S. Krishnan supervised the fabrication methodology. A. Prabharkar conceptualized and supervised the study and contributed to the writing process.

\bibliographystyle{styles/bibtex/splncs03_unsrt}
\bibliography{templates/references, templates/ref}

\end{document}